\documentclass[epj]{svjour}
\usepackage{graphics}
\begin{document}
\title{{\tt\bf CLAS12} and its Science Program at the Jefferson Lab Upgrade.} 
\subtitle{Selected Topics}
\author{Volker D. Burkert}
%
%
\institute{Jefferson Lab, Newport News, Virginia, USA}
\date\today
%
\abstract{An overview of the {\tt CLAS12} detector is presented and the initial physics program after the energy-doubling of the Jefferson Lab electron accelerator. Construction of the 12 GeV upgrade project has started October 2008. A broad program has been developed to map the nucleon's 3-dimensional spin and flavor content through the measurement of deeply exclusive and semi-inclusive processes. Other programs include forward distribution function to large $x_{B} \le 0.85$ and of the quark and gluon polarized distribution functions, and nucleon ground state and transition form factors at high $Q^2$. The 12 GeV electron beam and the large acceptance of CLAS12 are also well suited to explore hadronization properties using the nucleus as a laboratory.}      
%
\maketitle
\section{Introduction}
\label{intro}

The challenge of understanding nucleon electromagnetic structure still 
continues after more than five decades of experimental scrutiny. From the initial 
measurements of elastic form factors to the accurate determination of 
parton distributions through deep inelastic scattering (DIS), the
experiments have increased in statistical and systematic accuracy.  It was
realized in recent years that the parton distribution functions
represent special cases of a more general, and much more powerful  way of 
characterizing the structure of the nucleon, the generalized parton 
distributions (GPDs)~\cite{Ji:1996nm,Ji:1996ek,Radyushkin:1996nd,Radyushkin:1997ki}.

  The GPDs describe the
simultaneous distribution of particles with respect to both position and 
momentum. In addition to the information about the spatial density (form factors) 
and momentum density (parton distribution), these functions reveal the 
correlation of the spatial and momentum distributions, {\it i.e.} how the 
spatial shape of the nucleon changes when probing quarks of 
different wavelengths.

The concept of GPDs has led to completely new methods of ``spatial imaging''
of the nucleon, either in the form of two-dimensional tomographic images, or 
in the form of genuine 
three-dimensional images.  GPDs also allow us to 
quantify how the orbital motion of quarks in the nucleon contributes to the 
nucleon spin -- a question of crucial importance for our understanding of 
the ``mechanics'' underlying nucleon structure.  The spatial view of the 
nucleon enabled by the GPDs provides us with new ways to test dynamical 
models of nucleon structure. 

\begin{figure}[here]
\resizebox{0.5\textwidth}{!}{%
\includegraphics{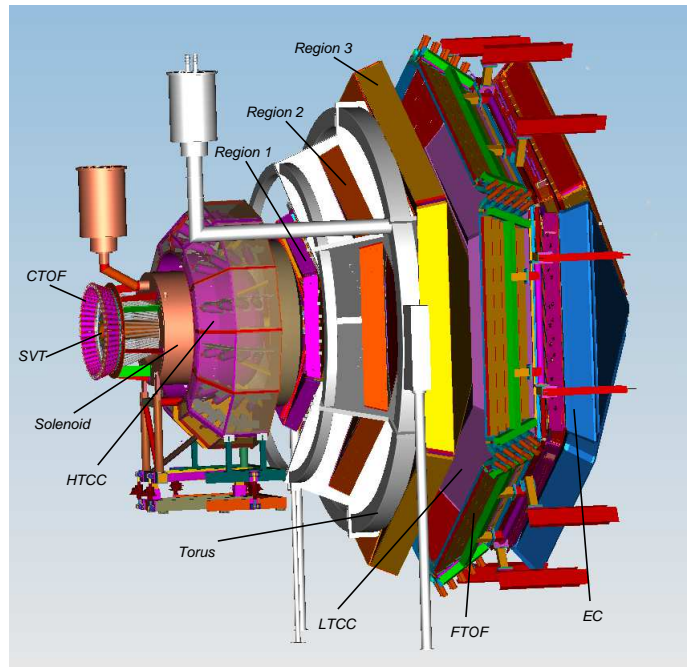}}
\caption{3D view of the CLAS12 detector. The beam comes from the left. The target is located inside the superconducting solenoid magnet.}
\label{CLAS12}    
\end{figure}

The mapping of the nucleon GPDs, and a detailed understanding of the
spatial quark and gluon structure of the nucleon, have been widely 
recognized as the key objectives of nuclear physics of the 
next decade. This requires a comprehensive program, combining results
of measurements of a variety of processes in electron--nucleon 
scattering with structural information obtained from theoretical studies, 
as well as with expected results from future lattice QCD simulations.

While GPDs, and also the recently introduced transverse momentum dependent 
distribution functions (TMDs), open up new avenues of research, the
traditional means of studying the nucleon structure through 
electromagnetic elastic and transition 
form factors, and through flavor- and spin-dependent parton distributions must also be 
employed with high precision to extract physics on the nucleon structure in the transition 
from the regime of quark confinement to the domain of asymptotic freedom. These 
avenues of research can be explored using the 12 GeV cw beam of the JLab 
upgrade with much higher precision than has been achieved before, 
and can help reveal some of the remaining secrets 
of QCD. Also, the high luminosity available 
will allow us to explore the regime of extreme 
quark momentum, where a single quark carries 80\% or more of the proton's 
total momentum.

\section{The {\tt CLAS12} Detector}
To meet the requirements of high statistics measurements for 
exclusive processes the equipment at JLab will undergo major upgrades. In particular it will include the {\tt CLAS12} 
large acceptance spectrometer~\cite{clas12_tdr}, which is shown in Fig.~\ref{CLAS12}. The main new features of {\tt CLAS12} include operation with a luminosity of $10^{35}$cm$^{-2}$sec$^{-1}$, 
an order of magnitude increase over previous CLAS~\cite{clas}, and improved particle identification capabilities at more forward angles. CLAS12 has two major parts with different functions, the Forward Detector (FD) and the Central Detector (CD). I this section I present a short descriptions of the detector system. 

\subsection{The Forward Detector} 

Improved electron-pion separation at higher momentum is achieved with 
a gas Cherenkov counter with a pion momentum threshold of 4.9 GeV/c. The new high threshold Cherenkov counter (HTCC) is positioned in front of a superconducting toroidal magnet, and has to present as little material to the charged particles as practical to limit multiple scattering contributing to the momentum resolution. This requires use of low mass composite material for the mirror system. 

The HTCC is followed by a toroidal magnet for the momentum analysis of tracks with scattering angles from $5^{\circ}$ to $40^{\circ}$. Similar to CLAS the new toroidal magnet has six superconducting symmetrically arranged around the beam line, and provides six sectors for charged particle detection.  In each sectors tracking is accomplished with a set of 3 regions of drift chambers with 12 layers of hexagoal drift cells arranged at stereo angles of $\pm 6^\circ$. This arrangement provides good angular resolution both in polar angle and in azimuthal angle. The drift chamber system will provide 36 measurements for a single charged track and has sufficient redundancy for pattern recognition and track reconstruction. 

The Torus magnet and the drift chamber system are followed by the low-threshold Cherenkov counter (LTCC) that provides charged pion identification for momenta greater than 3 GeV/c. Following the LTCC are two arrays of plastic scintillators for charged particle identification. The first layer contains 22 strips of 5cm thick and 15 cm wide scintillator strips and provides timing information of 150psec. The second layer is 6 cm thick and has 6 cm wide strips. It provides improved timing information of $\delta T < 100$ psec due to the better light collection. A combined resolution of~80 psec is expected. For equal pion, kaon, and proton yields this will enable a $4 \sigma$ $\pi$/K separation up to 3 GeV/c, and a K/p separation up to 4.5 GeV/c from time-of-flight measurements. A future upgrade of the particle identification is under consideration that would replace one or more of the LTCC sectors with RICH detectors that will allow much improved identification of pions, kaons and protons at high momentum where time-of-flight measurements are less effective.  

Large parts of the physics program require the identification of single high energy photons and separation from $\pi^\circ \rightarrow \gamma \gamma$ up to 9 GeV/c. The granularity of the existing electromagnetic calorimeter (EC) will be improved by adding a preshower calorimeter of 5-6 radiation length (PCAL) in front of EC that provides a factor 2.5 better spatial resolution and a separation of two photons up to momenta greater 10~GeV/c. At forward angles below 6$^\circ$, a lead-tungstate calorimeter consisting of 420 crystals with an average cross section of 15mm x 15mm and 18 rad. length thick, provides photon and $\pi^\circ$ identification for momenta up to 10~GeV/c.


\subsection{The Central Detector}

The Central Detectors (CD) is based on a compact solenoid magnet with maximum central magnetic field of 5 Tesla. The solenoid magnets provides momentum analysis for polar angles greater than 35$^\circ$, protection of tracking detectors from background electrons, and acts as a polarizing field for dynamically polarized solid state targets. All three functions require high magnetic field. The overall size of the solenoid is restricted to 200~cm in diameter which allows a maximum warm bore for the placement of detectors of 80~cm. To obtain sufficient momentum resolution in the limited space available requires high field and excellent position resolution of the tracking detectors. The central field in the target region must also be very uniform at $\Delta{B}/B < 10^{-4}$ to allow the operation of a dynamically polarized target. To achieve a sustained high polarization for polarized ammonia targets requires magnetic fields in excess of 3~T. Magnetic Fields of 5~T have been most recently used for such targets with polarization of 80\% - 90\% for hydrogen. In addition, the solenoidal field provides the ideal guiding field for keeping the copiously produced M\"oller electrons away from the sensitive detectors and guide them to downstream locations where they can be passively absorbed in high-Z material.        
     
Tracking in the CD is provided by a silicon vertex tracker (SVT) that uses silicon strip technology and provides tracking for polar angle from 5$^\circ$  to 135$^\circ$. The tracker consists of a barrel strip tracker (BST) and a forward detecor (FST). The BST has 8 stereo layers of silicon sensors and provides standalone tracking for polar angle from 35$^circ$  to 135$^\circ$. The FST has 6 stereo layers and covers the range from $5^\circ$ to $35^\circ$ and acts together with the forward drift chamber tracking system to significantly improve vertex resolution and momentum  resolution.   

The central time-of-flight scintillator array (CTOF) consists of 50 strips of fast plastic scintillator equipped with 100 photomultipliers that provide 2-sided light readout. The scintillator light is brought to an area of reduced magnetic field where either PMTs with lower magnetic field sensitivity can be used or where passive shielding can be employed. R\&D work is still underway to study these options. 

The very short flight path available allows for particle identification in a restricted momentum range of up to 1.2 GeV/c and 0.65 GeV/c for pion-proton and pion-kaon separation, respectively. 

Several possible upgrades of the CD are currently being studied: a significant improvement in tracking resolution is possible by adding several layers of micromegas detectors to the barrel part of the SVT, and replacing two of the SVT layers by micromegas. This option is currently under study at CEA Saclay. Also under study by a French-Italian collaboration is an additional detector that will add neutral particle detection capabilities, for example for the study of DVCS on neutrons, using a liquid deuterium target. This detector will fill the gap between the CTOF and the solenoid cryostat.    

With these upgrades {\tt CLAS12} will be the work horse for exclusive and semi-inclusive electroproduction experiments in the deep inelastic kinematics.


In the following sections I describe the currently anticipated initial physics program for CLAS12.

\begin{figure}[htb]
\resizebox{0.48\textwidth}{!}{%
  \includegraphics{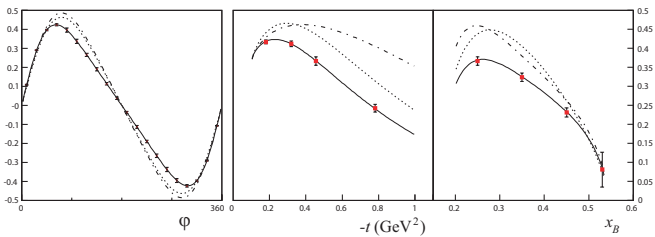}}
\caption{The beam spin asymmetry showing the DVCS-BH interference for 11 GeV beam 
energy~\cite{e12-06-119}. Left panel: $x=0.2$, 
$Q^2=3.3$GeV$^2$, $-t=0.45$GeV$^2$. Middle and right panels:  $\phi=90^{\circ}$, 
other parameters same as in left panel. Many other bins will be measured simultaneously. 
The curves represent various parameterizations within the VGG model~\cite{vgg}. Projected uncertainties are statistical.}
\label{fig:dvcs_alu_12gev}   
\end{figure}

\section{Generalized Parton Distributions and DVCS}
\label{sect:1}

It is now well recognized~\cite{Ji:1996nm,Belitsky:2001ns,Burkardt:2002hr,Belitsky2004} that 
deeply virtual Compton scattering and deeply virtual meson production are most suitable for
mapping out the twist-2 vector GPDs $H,~E$ and the axial GPDs ${\tilde H},~{\tilde E}$ in $x,~\xi,~t$, 
where $x$ is the momentum fraction of the 
struck quark, $\xi$ the longitudinal momentum transfer to the quark, and $t$ the 
transverse momentum transfer to the nucleon. Having access to a 3-dimensional image
of the nucleon (two dimensions in transverse space, one dimension in longitudinal 
momentum) opens up completely new insights into the complex structure of the
nucleon. In addition, GPDs carry information of more global nature. For example,
the nucleon matrix element of the energy-momentum tensor 
contains 3 form factors that encode information on the angular 
momentum distribution $J^q(t)$ of the quarks with flavor $q$ in transverse space, their 
mass-energy distribution $M_2^q(t)$, and their pressure and force 
distribution $d^q_1(t)$. How can we access these form factors? The only 
known process to directly measure them is elastic graviton scattering 
off the nucleon. However, these form factors also 
appear as moments of the vector GPDs~\cite{goeke2007}, thus offering
prospects of accessing gravitational form factors through the detailed mapping of GPDs.  The 
quark angular momentum in the nucleon is given by 
$$J^q(t) = 
\int_{-1}^{+1}dx x [H^q(x, \xi, t) + E^q(x, \xi, t)]~,$$ and the mass-energy and pressure distribution $$M_2^q(t) + 4/5d^q_1(t)\xi^2 
= \int_{-1}^{+1}dx x H^q(x, \xi, t)~.$$ The mass-energy and force-pressure distribution 
of the quarks are given by the second moment of GPD $\it{H}$, and their relative contribution is controlled by $\xi$. A separation of $M^q_2(t)$ and 
$d^q_1(t)$ requires measurement of these moments in a large range of 
$\xi$. The beam helicity-dependent cross section asymmetry is given 
in leading twist as 
$$ A_{LU} \approx \sin\phi[F_1(t)H + \xi(F_1+F_2)\tilde{H}]d\phi~, $$where
$\phi$ is the azimuthal angle between the electron scattering plane and the hadronic plane. The kinematically suppressed term with GPD $E$ is omitted. 
The asymmetry is mostly sensitive to the GPD $H(x=\xi,\xi,t)$. In a wide kinematics~\cite{clas-dvcs-1,clas-dvcs-3}  
the beam asymmetry $A_{LU}$ was measured at Jefferson Lab at modestly high $Q^2$, $\xi$, and $t$, and in a more limited kinematics~\cite{halla-dvcs} the cross section difference
$\Delta\sigma_{LU}$ was measured with high statistics. Moreover, 
a first measurement of the target asymmetry 
$A_{UL}=\Delta\sigma_{UL}/2\sigma$ was carried out~\cite{clas-dvcs-2}, where 
$$A_{UL} \approx \sin\phi[F_1\tilde{H} + \xi(F_1+F_2)H]~.$$  
The combination of $A_{LU}$ and $A_{UL}$ allows to separate GPD $H(x=\xi,\xi,t)$ and
$\tilde{H}(x=\xi,\xi,t)$.  
Using a transversely polarized target the asymmetry 
$$A_{UT} \approx \cos\phi\sin(\phi-\phi_s) [t/4M^2 (F_2H - F_1 E)] $$ can be measured, where $\phi_s$ is the azimuthal angle of the target polarization vector relative to the electron scattering plane. $A_{UT}$ 
depends in leading order on GPD $E$.

\begin{figure}
\resizebox{0.5\textwidth}{!}{%
  \includegraphics{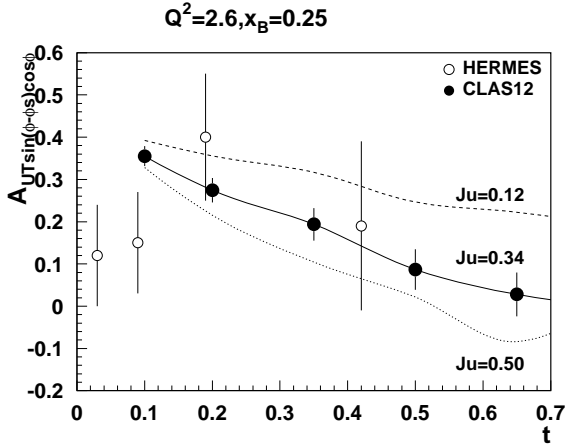}}
\caption{Projected transverse target asymmetry $A_{UT}$ for DVCS production off protons at 11 GeV beam energy.}
\label{fig:dvcs_aut_12gev}    
\end{figure}

\begin{figure}
\resizebox{0.5\textwidth}{!}{%
  \includegraphics{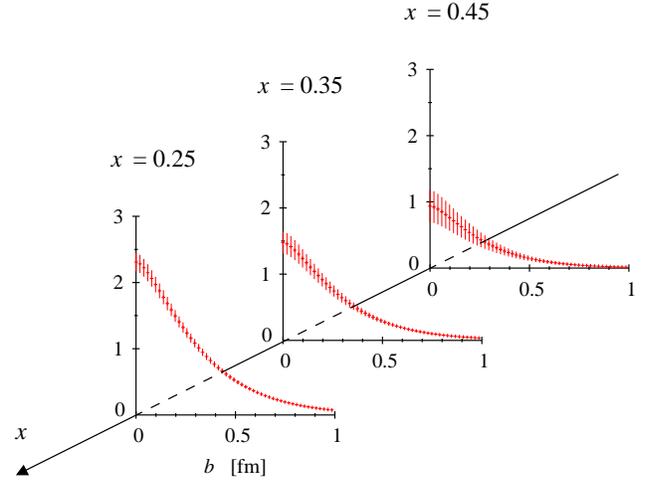}}
\caption{The u-quark distribution in transverse space as extracted from projected DVCS data with {\tt CLAS12}. }
\label{fig:gpd_H}    
\end{figure}

First DVCS experiments carried out at JLab~\cite{clas-dvcs-1,clas-dvcs-2,halla-dvcs,clas-dvcs-3} and DESY~\cite{hermes-dvcs} showed promising 
results in terms of the applicability of the handbag mechanism to probe GPDs. The 12 GeV upgrade offers much improved possibilities for accessing GPDs. 
Figure~\ref{fig:dvcs_alu_12gev} shows the expected statistical precision of 
the beam DVCS asymmetry for some sample kinematics. Using a
polarized target one can also measure 
the  target spin asymmetries with high precision. Figure~\ref{fig:dvcs_aut_12gev} shows the expected statistical accuracy for one kinematics bin. A measurement of all 3 
asymmetries will allow a separate determination of GPDs 
$H,~\tilde{H}$ and $E$ at the above specified kinematics. Through a Fourier transformation
the t-dependence of GPD $H$ can be used to determine the $u-$quark distribution 
in transverse impact parameter space. Figure~\ref{fig:gpd_H} shows projected 
results.

Deeply virtual meson production will play an important role in disentangling 
the flavor- and spin-dependence of GPDs. For exclusive mesons only the longitudinal 
photon coupling in $\gamma^* p \rightarrow Nm$ enables direct access to GPDs through 
the handbag mechanism and must be isolated from the transverse coupling. That the transverse
contribution cannot be neglected at currently available energies of 6 GeV was observed in 
the non-zero beam asymmetry measured with CLAS~\cite{demasi} that indicated the presence of 
a significant longitudinal-transverse interference term in the amplitudes. In addition, 
the dominance of the handbag mechanism in the longitudinal cross section must first be 
established at the upgrade energy.

\section{Transverse momentum dependent parton distributions and SIDIS}
\label{sect:2}

 Semi-inclusive deep inelastic scattering
(SIDIS) studies, when a hadron is detected in coincidence with the scattered 
lepton that allows ``flavor tagging'', provide more direct access to 
contributions from different quark species.  In addition, they give access to the 
transverse momentum distributions of quarks, not accessible in inclusive 
scattering.  Azimuthal distributions of final state particles in semi-inclusive 
deep inelastic scattering  provide access to the orbital motion of quarks and 
play an important role in the study of TMDs of 
quarks in the nucleon.
\begin{table}[h]
\begin{center}
\begin{tabular}{|c|c|c|c|} \hline\hline
N/q & U & L & T \\ \hline
 {U} & ${\bf f_1}$   & & ${ h_{1}^\perp}$ \\ \hline
 {L} & &${\bf g_1}$ &    ${ h_{1L}^\perp}$ \\ \hline
 {T} & ${ f_{1T}^\perp} $ &  ${ g_{1T}}$ &  ${ \bf h_1}$ \, ${ h_{1T}^\perp }$ \\
\hline\hline
\end{tabular}
\end{center}
\caption{\small{{Leading-twist transverse momentum-dependent distribution 
functions.  $U$, $L$, and $T$ stand for transitions of unpolarized, 
longitudinally polarized, and transversely polarized nucleons (rows) to 
corresponding quarks (columns).}}
\label{tab1}} 
\end{table}

\noindent TMD distributions (see Table~\ref{tab1}) describe transitions of a nucleon 
with one polarization in the initial state to a quark with another polarization 
in the final state. The diagonal elements of the table are the momentum, longitudinal and 
transverse spin distributions of partons, and represent well-known parton
distribution functions related to the square of the leading-twist, light-cone 
wave functions. Off-diagonal elements require non-zero orbital angular 
momentum and are related to the wave function overlap of $L$=0 and $L$=1 Fock 
states of the nucleon~\cite{Ji:2002xn}.  The chiral-even distributions 
$f_{1T}^\perp$ and $g_{1T}$ are the imaginary parts of the corresponding
interference terms, and the chiral-odd $h_1^\perp$ and $h_{1L}$ are the
real parts.  The TMDs $f_{1T}^\perp$ and  $h_{1}^\perp$, which are related to 
the imaginary part of the interference of wave functions for different orbital 
momentum states and are known as the Sivers and 
Boer-Mulders functions, and describe unpolarized quarks in the 
transversely polarized nucleon and transversely polarized quarks in the 
unpolarized nucleon respectively.  
The most simple mechanism that can lead to a Boer-Mulders function is a 
correlation between the spin of the 
quarks and their orbital angular momentum.  In combination with a final state 
interaction that is on average attractive, already a measurement of the sign 
of the Boer-Mulders function, would thus reveal the correlation between 
orbital angular momentum and spin of the quarks. 

\begin{figure}
\resizebox{0.5\textwidth}{!}{%
	\includegraphics{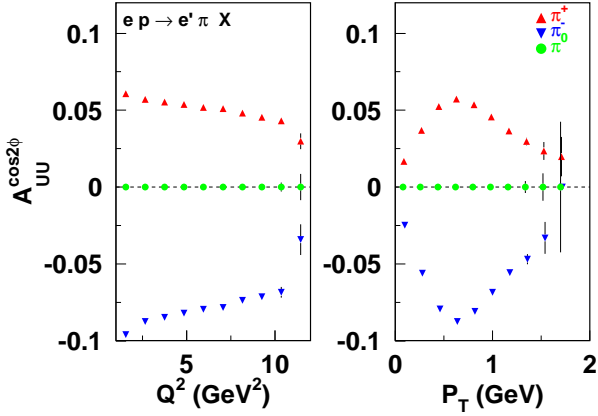}}
\caption{The $\cos2\phi$ moment (Boer-Mulders asymmetry) for pions
as a function of $Q^2$ and $P_T$ for $Q^2>2$~GeV$^2$ (right) with {\tt CLAS12} 
at 11~GeV from 2000~hours of running.  Values are calculated assuming
$H_1^{\perp u\rightarrow \pi^+}=-H_1^{\perp u\rightarrow \pi^-}$. Only statistical
uncertainties are shown.}
\label{fig:Boer-Mulders}    
\end{figure}

\begin{figure}[]
\resizebox{0.5\textwidth}{!}{%
	\includegraphics{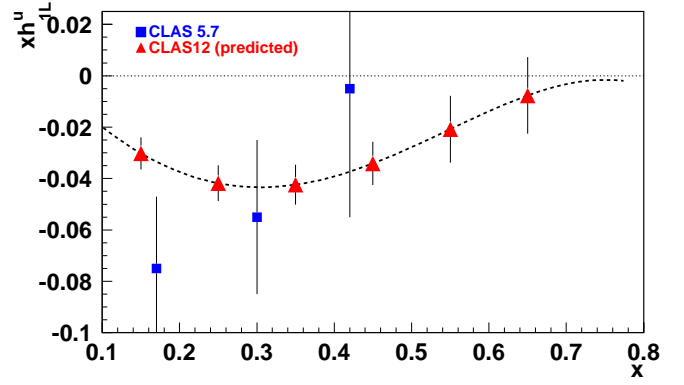}}
\caption{Projected data from {\tt CLAS12} for the chiral odd function $h^u_{1L}$ for $u$-quarks. }
\label{fig:h1lu11}    
\end{figure}

Similar to GPDs, TMD studies will benefit from the higher energy and high 
luminosity at 12 GeV. A comprehensive program is in preparation with {\tt CLAS12} to study the new structure functions.
Examples of expected uncertainties~\cite{E12-07-107} for the Boer-Mulders asymmetry $A^{cos2\phi}_{UU}$ 
are presented in Fig.~\ref{fig:Boer-Mulders}. Projections of the 
Mulders function $h^u_{1L}$ for $u-$quarks from $\pi^+$ asymmetries $A_{UL}$ with {\tt CLAS12} are shown in Fig.~\ref{fig:h1lu11}, and compared with preliminary results from the {\tt CLAS} EG1 data set at 5.75 GeV beam energy.

\section{Inclusive structure functions and moments}
\label{sect:3}
Polarized and unpolarized structure functions of the nucleon offer a
unique window into the internal quark structure of baryons.
The study of these structure functions provides insight into the two
defining features of QCD --- asymptotic freedom at small distances,
and confinement and non-perturbative effects at large distance scales.
After more than three decades of measurements at many accelerator
facilities worldwide, a truly impressive amount of data has been
collected, covering several orders of magnitude in both kinematic
variables $x$ and $Q^2$.
However, there are still important regions of the accessible phase space
where data are scarce and have large errors and where significant
improvements are possible through precise experiments at Jefferson Lab. 

\begin{figure}[bmt]
\resizebox{0.5\textwidth}{!}{%
  \includegraphics{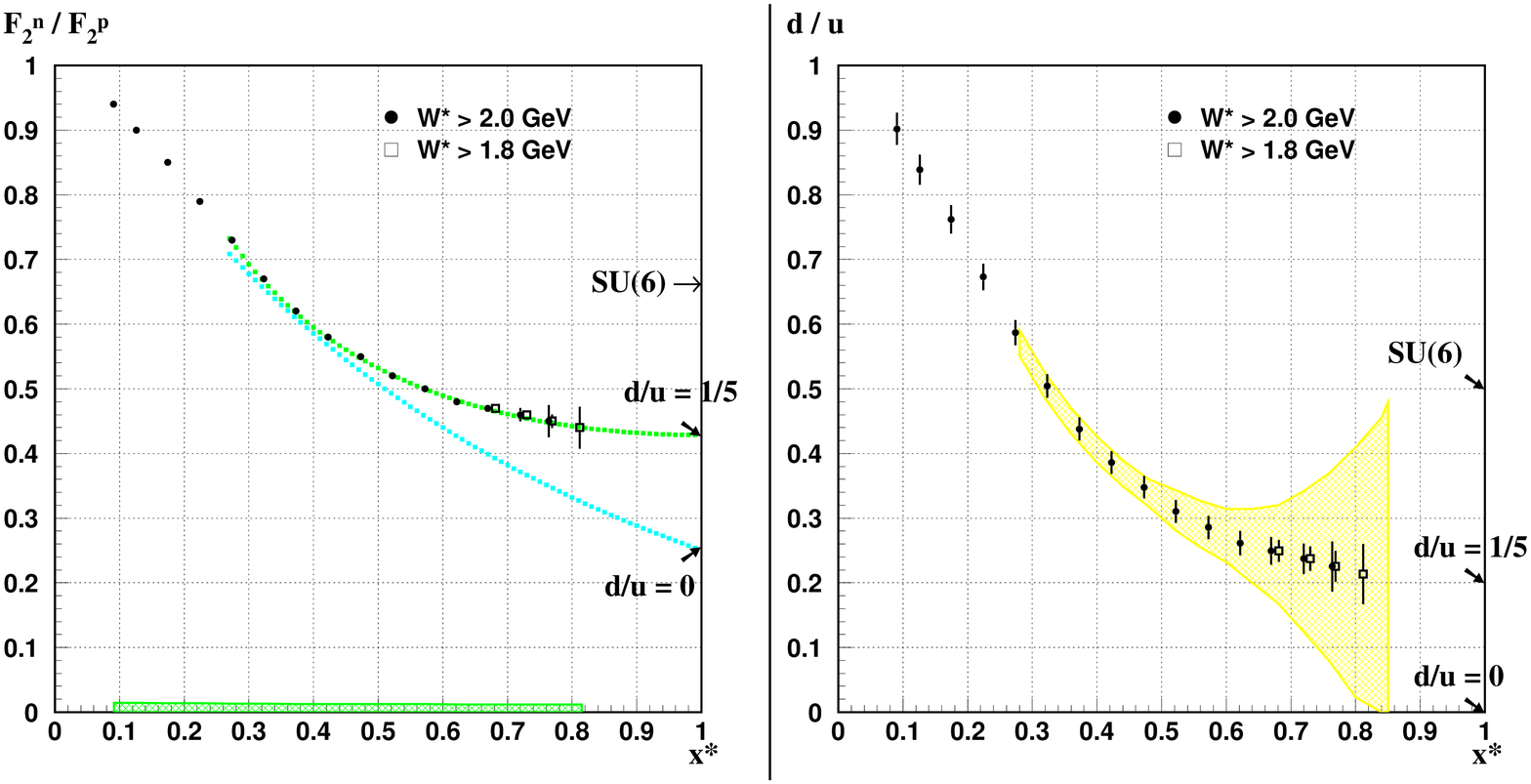}}
\caption{Projected data for the ratio $F^n_2/F^p_2$ (left) and $d/u$ (right) for 
11 GeV beam energy~\cite{bonus12_proposal}. The error bars in the right panel 
contain both statistical and 
systematic uncertainties. The yellow area shows the uncertainty of current data due to
poorly known nuclear corrections.}
\label{fig:F2nF2p}    
\end{figure}

One of the open questions is the behavior of the
structure functions in the extreme kinematic limit $x \rightarrow 1$.
In this region effects from the virtual sea of quark-antiquark pairs are 
suppressed, making this region simpler to model. This is also the region
where pQCD can make absolute predictions. However, the large $x$ 
domain is hard to reach because cross sections are kinematically suppressed, 
the parton distributions are small and final states interactions (partonic or 
hadronic) are large. First steps into the large $x$ domain became possible at energies
of 5-6 GeV \cite{BONUS,zheng2004,vipuli2006,bosted2007}. The interest triggered by these
first results and the clear necessity to extend the program to larger $x$ 
provided one of the cornerstone of the JLab 12 GeV upgrade physics program.  

\begin{figure}
\resizebox{0.5\textwidth}{!}{%
  \includegraphics{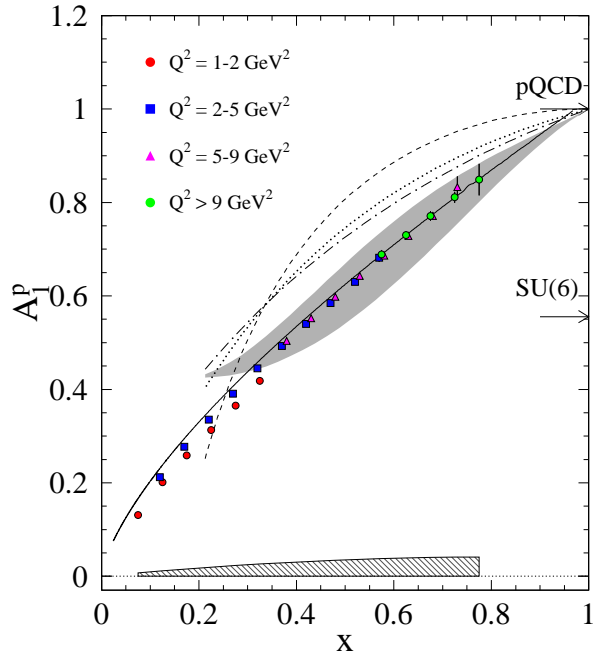}}
\caption{Anticipated results on $A_1^p$.The four different symbols represent four different $Q^2$ ranges.  The statistical uncertainty is given by the error bars while the systematic uncertainty is given by the shaded band. }
\label{fig:A1p}    
\end{figure}

\subsection{Valence quark structure and flavor dependence at large $x$.}
\label{sect:4}

The unpolarized structure function $F^p_2(x)$ has been mapped out in a 
large range of $x$ leading to precise knowledge of the quark distribution 
$u(x)$.  The corresponding structure function $F^n_2(x)$ is, however, well 
measured only for $x < 0.5$ as nuclear corrections, when using deuterium as 
a target, become large at large $x$ and are not well represented by 
Fermi motion. 
At JLab a new technique tested recently with {\tt CLAS} has been shown to be very
effective in reducing the nuclear corrections. The 
{\tt BONUS} experiment~\cite{BONUS} 
has recently taken data using a novel radial TPC with GEM readout as detector for 
the low-energy spectator proton in the reaction $en(p_s)\rightarrow ep_sX$.
Measurement of the spectator proton for momenta as low as 70 MeV/c and at large 
angles minimizes the poorly known nuclear corrections at large $x$. The 
techniques can be used with {\tt CLAS12} at 12~GeV to accurately determine 
the ratio $d(x)/u(x)$ to much larger $x$ values. 
Figure~\ref{fig:F2nF2p} shows the projected data for $F^n_2(x)/F^p_2(x)$ and $d(x)/u(x)$. 
A dramatic improvement can be achieved at large $x$.

\subsection{Spin structure functions and parton distributions}
\label{sect:5}

\begin{figure}
\resizebox{0.5\textwidth}{!}{%
  \includegraphics{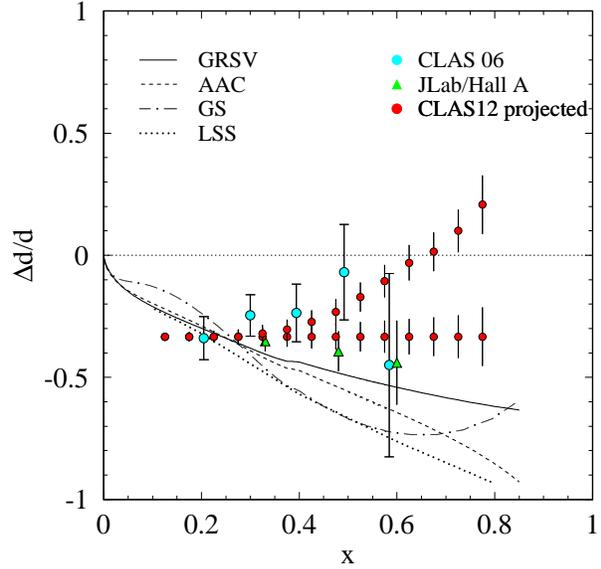}}
\caption{Expected results for $(\Delta d+ \Delta\bar{d})/(d+\bar{d})$. The central values of the data are following two arbitrary curves to demonstrate how the two categories of predictions, namely the ones that predict $\Delta d/d$ stays negative (LO and NLO analyses of polarized DIS data: GRSV, LSS, AAC, GS, statistical model, and a quark-hadron duality scenario) and the ones predicting $\Delta d/d \to 1$ when $x \to 1$ (leading order pQCD and a quark-hadron duality scenario).}
\label{fig:ddodxpctd}    
\end{figure}

The JLab PAC30 also approved E12-06-109 \cite{EG12} which will specifically study polarized parton distributions at large $x$. Using standard detection equipment, a redesigned polarized target adapted to {\tt CLAS12} and 30 (50)~days of running on a longitudinally polarized NH$_3$ (ND$_3$) target, high precision measurements can be achieved as shown in Fig.~\ref{fig:A1p}. These data will disentangle models in the large-$x$ region.  While the results shown in Fig.~\ref{fig:A1p} are with a $W>2$~GeV constraint, hadron-parton duality studies will tell us by how much this constraint can be relaxed, possibly increasing the $x$ range up to 0.9.  The expected accuracy for $(\Delta d+ \Delta\bar{d})/(d+\bar{d})$ is shown in Fig.~\ref{fig:ddodxpctd}.
  
\subsection{Global Fit of Polarized Parton Distributions}
\label{sect:6}
\begin{figure}
\resizebox{0.5\textwidth}{!}{%
  \includegraphics{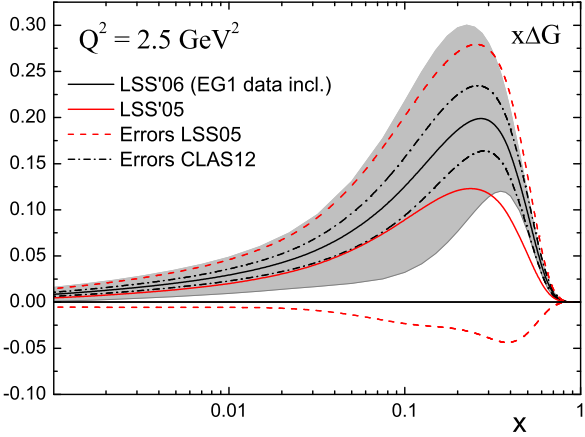}}
\caption{Expected uncertainties for $x\Delta{G}$. The black solid curve shows the central value of the present analysis that includes CLAS EG1 data. The dashed-dotted lines give the error band when the expected {\tt CLAS12} data are included in the LSS QCD analysis.}
\label{fig:pPDFs_exp}    
\end{figure}

The large window that wil open up over the DIS domain with the 12 GeV upgrade will permit 
constraints of global fits of the parton distributions.  JLab data at lower 
energies had already unique impact
at large $x$. The improvement from the 12-GeV upgrade is 
also significant at low and moderate $x$, noticeably for the polarized gluon 
distribution $\Delta G$.  To demonstrate the precision 
achievable with the expected {\tt CLAS12} data, we have plotted in 
Fig.~\ref{fig:pPDFs_exp} a study of the expected impact of expected JLab data on the 
NLO analyses of the polarized gluon distribution~\cite{LSS2007}.  A dramatic 
improvement can be achieved with the expected data from the {\tt CLAS12} 
proposal E12-06-109~\cite{EG12}.  We emphasize that the data will not only 
reduce the error band on $\Delta G$, but will likely allow a more detailed 
modeling of its $x$-dependence. Significant improvements are expected for the quark 
distributions as well.

\subsection{Moments of spin structure functions.}
\label{sect:7}
Moments of structure functions provide powerful insight into nucleon 
structure.  Inclusive data at JLab have permitted evaluation of the moments 
at low and intermediate $Q^2$~\cite{fatemi2003,yun2003,chen2004}.  
With a maximum beam energy of 6~GeV, however, the measured strength of the 
moments becomes rather limited for $Q^2$ greater than a few GeV$^2$. The 
12-GeV upgrade removes this problem and allows for measurements to higher 
$Q^2$. 

\begin{figure}
\resizebox{0.5\textwidth}{!}{%
  \includegraphics{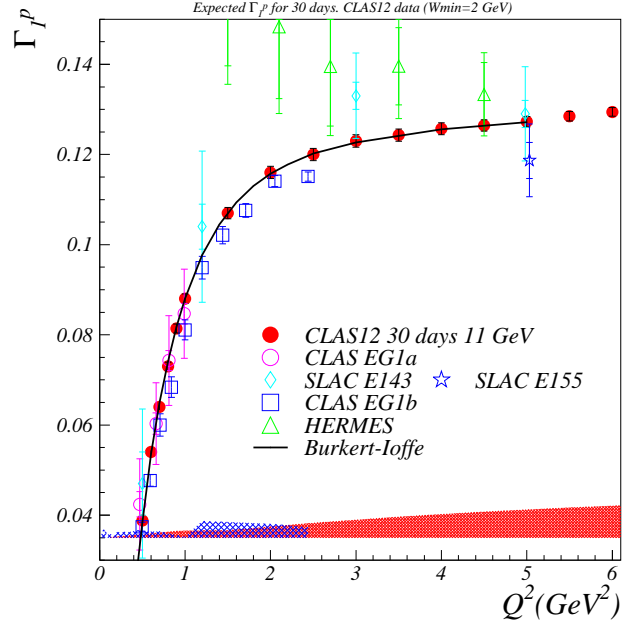}}
\caption{Left plot: expected precision on $\Gamma_1^p$ for {\tt CLAS12}
and 30~days of running.  {\tt CLAS} EG1a~\cite{fatemi2003,yun2003} data 
and preliminary results from EG1b are shown for comparison.  The data and 
systematic uncertainties include estimates of the unmeasured DIS 
contribution.  HERMES~\cite{Airapetian:2002wd} data, and E143~\cite{Abe:1998wq} 
and E155 data~\cite{Anthony:2000fn} from SLAC are also shown (including DIS 
contribution estimates).  The model is from Burkert and Ioffe
\cite{Burkert:1992tg,Burkert:1993ya}.}
\label{fig:expect}    
\end{figure}

Moments of structure functions are related to the nucleon static properties 
by sum rules. At large $Q^2$ the Bjorken sum rule relates the integral $\Gamma_1^{p-n}=\int(g_1^p- g_1^n)dx$ 
to the nucleon axial charge~\cite{Bjorken:1966jh}. 
Figure~\ref{fig:expect} shows the expected precision on $\Gamma_1^p$. Published 
results and preliminary results from EG1b are also displayed for comparison. 
The hatched blue band corresponds to the systematic uncertainty 
on the EG1b data points.  The red band indicates the estimated systematic 
uncertainty from {\tt CLAS12}.  The systematic uncertainties for EG1 and 
{\tt CLAS12} include the estimated uncertainty on the unmeasured DIS part 
estimated using the model from Bianchi and Thomas~\cite{Thomas:2000pf}.  As 
can be seen, moments can be measured up to $Q^2$=6~GeV$^2$ with a statistical 
accuracy improved several fold over that of the existing world data.

Finally, moments in the low ($\simeq$ 0.5 GeV$^2$) to moderate 
($\simeq$4~GeV$^2$) $Q^2$ range enable us to extract higher-twist parameters,
which represent correlations between quarks in the nucleon.  This extraction 
can be done by studying the $Q^2$ evolution of first moments~\cite{Osipenko2005,Chen:2005td}.
Higher twists have been consistently found to have, overall, a surprisingly 
smaller effect than expected.  Going to lower $Q^2$ enhances the higher-twist 
effects but makes it harder to disentangle a high twist from the yet higher 
ones.  Furthermore, the uncertainty on $\alpha _s$ becomes prohibitive at low 
$Q^2$.  Hence, higher twists turn out to be hard to measure, even at the 
present JLab energies.  Adding higher $Q^2$ to the present JLab data set 
removes the issues of disentangling higher twists from each other and of the 
$\alpha _s$ uncertainty.  The smallness of higher twists, however, requires 
statistically precise measurements with small point-to-point correlated 
systematic uncertainties.  Such precision at moderate $Q^2$ has not been 
achieved by the experiments done at high energy accelerators, while JLab at 
12~GeV presents the opportunity to reach it considering the expected 
statistical and systematic uncertainties of E12-06-109.  The total 
point-to-point uncorrelated uncertainty on the twist-4 term for the Bjorken 
sum, $f_2^{p-n}$, decreases by a factor of 5-6 compared to results obtained in
Ref.~\cite{Deur:2004ti}. 
%
\section{Nucleon form factors and resonance transitions at short distances}
\label{sect:8}
  
The most basic observables that reflect the composite nature of the 
nucleon are its electromagnetive form factors. Historically the first 
direct indication that the nucleon is not elementary comes from 
measurements of these quantities in elastic $ep$ scattering~\cite{HOF}. 
The electric and magnetic form factors characterize the distributions of 
charge and magnetization in the nucleon as a function of spatial resolving 
power. The transition form factors reveal the nature of the excited states of the 
nucleon.  Further, these quantities can be described and related to other 
observables through the GPDs.

\begin{figure}[t]
\resizebox{0.45\textwidth}{!}{%
  \includegraphics{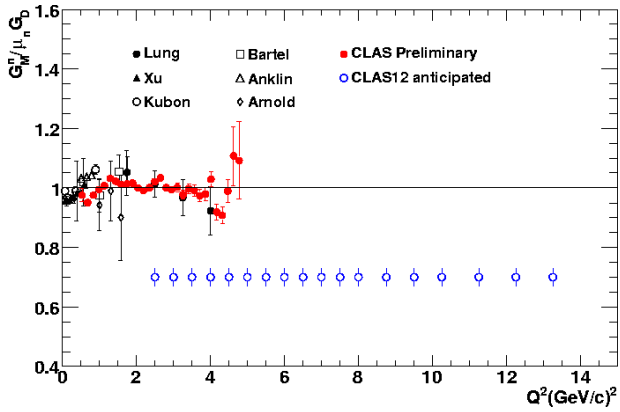}}
\caption{The magnetic form factor for the neutron. The existing data, and projected uncertainties at 12 GeV with {\tt CLAS12} (blue open circles).}
\label{fig:gmn}       
\end{figure}

Measurements of the elastic form factors will remain an important aspect 
of the physics program at 12 GeV, and will be part of the program in other 
experiments at JLab. The magnetic form factor of the 
neutron, as well as the transition form factors  for several prominent 
resonances require special experimental setups for which {\tt CLAS12} 
is suited best. Figure ~\ref{fig:gmn} shows the current data as well as the extension 
in $Q^2$ projected for the 12 GeV program with {\tt CLAS12}. 

\begin{figure}
\resizebox{0.4\textwidth}{!}{%
  \includegraphics{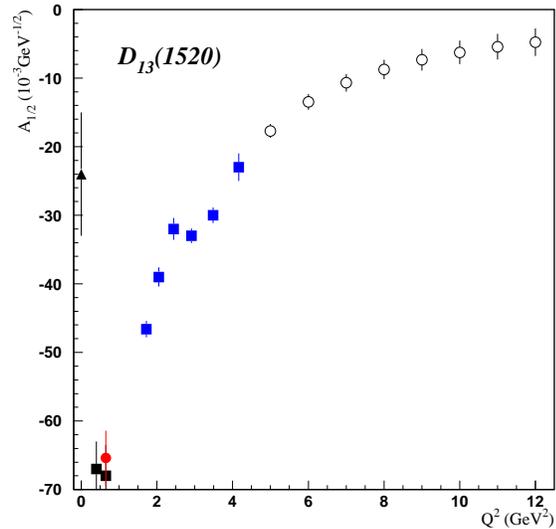}}
\caption{The transverse photocoupling amplitude $A_{1/2}$ for the 
$N(1520)D_{13}$ resonance.
The blue full squares are preliminary data from CLAS. The open 
circles represent projected results with {\tt CLAS12} at 12 GeV.}
\label{fig:d13_high}       
\end{figure}

Nucleon ground and excited states represent different eigenstates of the Hamiltonian,
therefore to understand the interactions underlying nucleon formation from fundamental 
constituents, the structure of both the ground state and the excited states must be studied. The current 
$N^*$ program at Jlab has already generated results for the transition 
form factors at $Q^2$ up to 6 GeV$^2$ for the $\Delta(1232)$~\cite{frolov1999,joo2002,ungaro2006}, and up 
to 4~GeV$^2$ for the $N(1535)S_{11}$~\cite{armstrong1999,thompson2001,denizli2007}. The most 
recent results~\cite{park2007,aznauryan2007}  
on the transition form factors for the Roper resonance $N(1440)P_{11}$ for $Q^2$ up to 4.5~GeV$^2$, have 
demonstrated the sensitivity to the degrees 
of freedom that are effective in the excitation of particular states. 
The JLab energy upgrade will allow us to probe 
resonance excitations at much higher $Q^2$, where the relevance of elementary 
quarks in the resonance formation may become evident through the approach
to asymptotic scaling. Figure~\ref{fig:d13_high} shows projected $Q^2$ dependence 
of the $A_{1/2}$ transition amplitude for the $N(1520)D_{13}$ resonance obtained from single pion production. 
Higher mass resonances may be efficiently measured in double-pion processes~\cite{ripani2003,mokeev} 
such as $ep\rightarrow ep\pi^+\pi^-$.    

\section{Quarks and hadrons in the nuclear medium}

\subsection{Color Transparency}

\begin{figure}[t]
\resizebox{0.4\textwidth}{!}{%
  \includegraphics{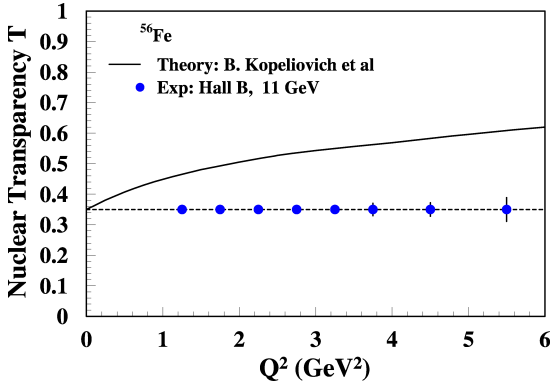}}
\caption{Projected color transparency effects in Fe. The open 
circles represent projected results with {\tt CLAS12} at 12 GeV.}
\label{CLAS12_ct}   
\end{figure}

Color transparency (CT) is a unique prediction of QCD, and implies that under the right conditions, 
the nuclear medium will allow the transmission of hadrons with reduced absorption. The phenomenon 
of CT is predicted on the quark-gluon basis and is totally unexpected on a hadronic interaction 
picture. Three ingredients are necessary to observe CT: (1) the interactions 
must create a small size object (point-like configuration, PLC) that (2) has a small cross section 
when traveling in a hadronic medium, and (3) the distance over which it expands to its full hadronic 
size must be larger than the nucleus size.    
Such conditions require high enough energy transfer to the target where the photon couples to PLCs, 
and the full hadronization occurs outside the nucleus. The energy doubling of the electron accelerator 
to 12 GeV will provide conditions where a significantly increased transparency should be observable. 
Small increases in nuclear transparency consistent with theoretical predictions have been observed at 
JLab with 5-6 GeV electron beams in pion production~\cite{ct_hallc}, and in $\rho^\circ$ electroproduction with CLAS~\cite{ct_clas}. At 12 GeV much more significant changes of nuclear transparency are predicted and can be observed with high sensitivity in CLAS12 as shown in Fig.~\ref{CLAS12_ct}.

\subsection{Hadronization} 
The use of electron beams at 12 GeV allows us to address fundamental questions of how colored quarks struck in the interaction with high energy photons transform into colorless hadrons. Questions that we want to have answered are, how long can a light colored quark remain deconfined? The production time $T_p$ measured this quantity. Since deconfined quarks emit gluons, $T_p$ can be measured via medium-stimulated gluon emission resulting in a broadening of the transverse momentum distribution of the final hadrons. Another important question to address is: How long does it take to form the color field of a hadron? This can be measured by the formation time $T_f^h$. Since hadrons interact strongly with the nuclear medium, $T_f^h$ can be determined by measuring the attenuation of hadrons in the nuclear medium by using nuclei of different sizes.

These question can be addressed by measuring the hadronic multiplicity ratio 
$$R_M^h(z,\nu,p_T^2,Q^2,\phi) = {\{{N_h^{DIS}(z,\nu,p^2_T,Q^2,\phi) \over N_e^{DIS}(\nu,Q^2)}\}_A \over  \{{N_h^{DIS}(z,\nu,p^2_T,Q^2,\phi) \over N_e^{DIS}(\nu,Q^2)}\}_D }$$ 
versus all kinematical quantities.  

\section{Conclusions}
\label{summary}

The JLab energy upgrade and the planned new experimental equipment are well 
matched to an exciting scientific program aimed at studies 
of the complex nucleon structure in terms of the newly discovered 
longitudinal and transverse momentum dependent parton distribution 
functions, the GPDs and TMDs. 
They provide fundamentally new insights in the complex multi-dimensional 
structure of the nucleon. In addition, the high precision afforded by the 
high luminosity and the large acceptance detectors, and the development of 
novel techniques to measure scattering off nearly free neutrons, will enable 
the exploration of phase space domains with extreme conditions that could not be 
studied before. The {\tt CLAS12} detector will play a crucial role in exciting program.

\vspace{0.3cm}

\noindent{\bf Acknowledgment}

I am grateful to members of the CLAS collaboration who contributed to the development of the exciting physics program for the JLab upgrade to 12 GeV, and the {\tt CLAS12} detector. Much of the material in this report is taken from the {\tt CLAS12} Technical Design Report Version 3, October 2007~\cite{clas12_tdr}.  

This work was supported in part by the U.S. Department of Energy and the National Science Foundation, the French Commisariat {\'{a}} l'Energie Atomique, the Italian Instituto Nazionale di Fisica Nucleare, the Korea Research Foundation,  and a research grant of the Russian Federation. The Jefferson Science Associates, LLC, operates Jefferson Lab under contract DE-AC05-060R23177.


\begin{thebibliography}{}
\bibitem{clas12_tdr} The complete document may be obtained from the author.
\end{thebibliography}
\end{document}